\documentclass[12pt,letter]{article}
\usepackage{amsmath,latexsym,amssymb,amsfonts,amsbsy,mathrsfs,bm,braket,amscd,color}
\usepackage[mathscr]{eucal}
\usepackage{graphicx,multirow,latexsym}
\usepackage{amsfonts,delarray,float}

\usepackage[comma]{natbib}
\usepackage{url}

\usepackage{algorithm}
\usepackage{algorithmic}

\usepackage{fancyhdr}
\usepackage[tight,footnotesize]{subfigure}

\numberwithin{equation}{section}

\DeclareMathOperator*{\argmax}{arg\,max}
\DeclareMathOperator*{\argmin}{arg\,min}

\newcommand{\mL}{\mathcal L}
\newcommand{\mS}{\mathcal S}
\newcommand{\mP}{\mathcal P}

\newtheorem{thm}{Theorem}[section]
\newtheorem{cor}{Corollary}[section]

\newtheorem{exmp}{Example}[section]

\newcommand{\blind}{0}
\renewcommand{\baselinestretch}{1.5}

\begin{document}

\if0\blind
{
  \title{\bf Sensible Functional Linear Discriminant Analysis}
  \author{Lu-Hung Chen\footnote{Lu-Hung Chen is Assistant Professor, Institute of Statistics, National Chung-Hsing University, Taichung 402, Taiwan. Email: luhung@nchu.edu.tw.}
    and 
    Ci-Ren Jiang\footnote{Ci-Ren Jiang is Associate Research Fellow, Institute of Statistical Science, Academia Sinica, Taipei 115, Taiwan. Email: cirenjiang@stat.sinica.edu.tw.}} 
    
    \date{}
  \maketitle
} \fi
\def\spacingset#1{\renewcommand{\baselinestretch}%
{#1}\small\normalsize} \spacingset{1}

\begin{abstract}
The focus of this paper is to extend Fisher's linear discriminant analysis (LDA) to both densely re-corded functional data and sparsely observed longitudinal data for general $c$-category classification problems. We propose an efficient approach to identify the optimal LDA projections in addition to managing the noninvertibility issue of the covariance operator emerging from this extension. A conditional expectation technique is employed to tackle the challenge of projecting sparse data to the LDA directions. We study the asymptotic properties of the proposed estimators and show that asymptotically perfect classification can be achieved in certain circumstances. The performance of this new approach is further demonstrated with numerical examples. 

\end{abstract}


\section{Introduction}
Classification identifies the class, from a set of classes, to which a new observation belongs, based on the training data containing observations whose class labels are known. Due to its importance in many applications, statistical approaches have been extensively developed. To name but a few, principal component analysis (PCA, \cite{TurkP:91}), Fisher's linear discriminant analysis (LDA, \cite{Fisher:36,Rao:48}), partial least square approaches (PLS, \cite{BarkerR:03}), etc. have all been explored for classification. The common essence of these approaches is to find \emph{optimal projections} based on a particular criterion for subsequent classification. While the data dimension is moderate, these approaches or their variants often work nicely. With the advent of modern technology and devices for collecting data, the dimension of data can become very high and may be intrinsically infinite, such as functional data; this requires the aforementioned approaches to be adapted. Motivated by the Fisher's LDA, we propose ``sensible'' functional LDA (sFLDA) to search the \emph{optimal projections} for subsequent classification.

LDA aims at finding ideal linear projections and performs classification on the projected subspace. Ideal projections are those maximizing the projected distances between classes while keeping the projected distances among subjects in the same class minimized. Take a $p$-dimensional case for example; mathematically the ideal projections are the eigenvector $\bm b$ in 
\begin{equation}\label{Fisherdir}
\Sigma_W^{-1}\Sigma_B \bm b=\lambda \bm b,
\end{equation}
where $\Sigma_W^{-1}$ denotes the inverse of the within-subject covariance matrix $\Sigma_W$, and $\Sigma_{B}$ is the between covariance matrix that characterizes the variation of class means.  Under classical multivariate settings, $\Sigma_W$ is invertible. Please refer to \cite{MardiaKB:80} for the details of LDA. Due to its simplicity, LDA has been widely employed in many applications. 

Extending (\ref{Fisherdir}) directly to functional data is tricky due to the noninvertible covariance operator. Specifically, the inverse of the covariance operator is unbounded if the functional data is in $\mL_2$, which is commonly assumed in the functional data analysis literature (e.g., \cite{HallMW:06,LiH:10:1,DelaigleH:12}, etc.). To elucidate our idea, let us introduce notations first. Suppose the data consists of $c$ classes. Let $X_k$ be an $\mL_2$ stochastic process, defined on a finite compact interval $\mathcal{T}$, in class $k$ with mean function $\mu_k$ and a common covariance function $\Gamma_W$. Mercer's theorem implies that the covariance function can be further decomposed as $\Gamma_W(s,t) = \sum_{j=1}^\infty \lambda_j \phi_j(s) \phi_j(t),$ where the eigenvalue $\lambda_j > 0$ is in descending order with corresponding eigenfunction $\phi_j$ and $\sum_{j=1}^\infty \lambda_j < \infty$. Functional principal component analysis (FPCA) corresponds to a spectral decomposition of the covariance and leads to the well-known Karhunen-Lo\`{e}ve decomposition of the random function, 
\begin{equation}\label{eq:Xkkl}
X_k(t) = \mu_k(t)+\sum_{j=1}^\infty A_{k,j} \phi_j(t),
\end{equation}
where $A_{k,j}$ is the $j$th principal component score (PCS) with mean zero, variance $\lambda_j$ and $t\in\mathcal{T}$. Let $\mS_B$ (resp. $\mS_W$) be the space spanned by $\{\mu_k\}_{k=1}^c$ (resp. $\{\phi_j\}_{j=1}^\infty$). Since we do not assume completeness on $\{\phi_j\}_{j=1}^\infty$, $\mS_B \subset \mS_W$ is not always true \citep{HsinE:15}. We also do not impose any parametric assumptions on $X_k$ other than smoothness conditions on $\mu_k$ and $\Gamma_W$, which are quite common in functional data analysis (e.g., \cite{RiceS:91,ChioMW:03,HallMW:06}, etc.)  %

To handle the unbounded $\Gamma_W^{-1}$, basis-based approaches can be used to express the functional data with certain basis functions and turn the functional problem into a multivariate one. For example, \citet{HallPP01}, \citet{GleH:03}, \citet{Muller:05}, \citet{LengM:06}, and \citet{SongDLK:08} performed classification based on FPCA; \citet{PredaSL:07} classified functional data by means of PLS; \citet{BerlinetBR:08}, \citet{RinconRM:12}, and \citet{ChangCO:14} developed approaches based on wavelets. However, doing so might lose crucial information for subsequent classification if the differences among classes are not well preserved due to inappropriate basis functions. For example, when $\mS_B \nsubseteq \mS_W$ (e.g., a binary case where $\mu_1(t) = \sin(2\pi t)$, $\mu_2(t)=-\mu_1(t)$, $\phi_k(t) = \sqrt{2}\cos(2k\pi t)$ for $k=1,\ldots,\infty$ and $t\in [0,1]$), at least some $\mu_k$'s can not be well described by $\{\phi_j\}_{j=1}^\infty$ and thus FPCA based approaches might not be a good choice. This argument is substantiated with simulated data in section \ref{sec:sim}. 

There exist other functional classification approaches under different considerations. To name a few, \citet{FerrV:03} and \citet{GaleanoaJL:14} investigated distance-based approaches, \citet{HastieBT:95} and \citet{ArakiKKM:09} developed regularized approaches, \citet{Epifanio:08} proposed an approach to classify functional shapes, and \cite{DelaH:13} developed a functional classification framework when the observations were fragments of curves.

In a general $c$-category classification problem, at most ($c-1$) projections in $\mS_B$ are useful for functional LDA. Merely considering the information in $\mS_B$ is insufficient, as reducing the within-class variation is equivalently important. With this in mind and to properly handle the noninvertibility issue of $\Gamma_W$, we propose a sensible classification approach to find projections in $\mP_W^{\perp}(\mS_B)$ and in $\mP_W(\mS_B)$ sequentially, where $\mP_W(\mS_B)$ (resp. $\mP_W^{\perp}(\mS_B)$) is the projection of $\mS_B$ on $\mS_W$ (resp. $\mS_W^\perp$, the orthogonal complement of $\mS_W$). Most existing approaches do not appear to appreciate that the optimal linear projections could be a set of the projections obtained in $\mP_W^{\perp}(\mS_B)$ and in $\mP_W(\mS_B)$; this may be because it suffices to consider projections in either $\mP_W^{\perp}(\mS_B)$ or $\mP_W(\mS_B)$ for binary classification problems. Accordingly, our procedure is more general. 
   
Despite the difference in sampling schemes, functional data and longitudinal data come from similar sources. Therefore, it is practical to develop unified approaches for them (e.g., \cite{Muller:05,HallMW:06,JianW:10}, etc.). \cite{JamesH:01} employed natural cubic splines to tackle the problem of sparsity. \cite{WuL:13} applied the FPCA approach proposed in \cite{YaoMW:05:1} to reconstruct sparsely observed longitudinal data and performed robust support vector machine (SVM) on the reconstructed curves. This strategy leads to the same predicament as other FPCA based approaches mentioned earlier. The major challenge in extending Fisher's LDA to longitudinal data is to perform classification on a new subject with longitudinal observations. The sparsity and irregularity of the observations make the projections difficult. We propose an imputation approach based on a conditional expectation technique (in section \ref{reconstruct}) to resolve the sparsity issue without losing the subtle information about the mean functions. 

The rest of this paper proceeds as follows. In the next section, the motivation and the framework of sFLDA are introduced. The proposed estimators and their asymptotic properties are provided in sections 3 and 4, respectively. We propose an imputation approach for longitudinal data while performing projections in section 5. In section 6, simulation studies under three data configurations are conducted. In section 7, our approach, along with some competitors, is applied to two real data examples. Conclusions are given in the last section. Appendices include the assumptions made for the asymptotics, the leave-one-curve-out cross-validation (CV) formulae of bandwidth selections, and some details for Section 2.1. All the proofs are contained in the supplementary material.  

\section{Method} \label{method}


Let us elucidate our idea through the following example, where we aim to find the optimal projections, $\beta$'s, for subsequent classification. For convenience, we denote $\langle \beta, X \rangle = \int_\mathcal{T} \beta(t)X(t)dt$. 
\begin{exmp}\label{exmp1}
Suppose the data in class $k$ is generated from 
\begin{equation*}
X_{k,i}(t)  = \mu_k(t)+\sum_{j=1}^{\infty}A_{k,i,j} \phi_j(t),
\end{equation*}
where $\phi_{j}(t)=\sqrt{2}\sin(2\pi jt)$, $t\in[0,1]$, $A_{k,i,j}\sim N(0,1/j^2)$, and $i=1,\ldots,n_k$. 
\begin{align*}
\text{Case (a): }& \mu_1(t)=\sqrt{2}\sin(2\pi t), \mu_2(t)=0;\\
\text{Case (b): }& \mu_1(t)=\sqrt{2}\cos(2\pi t), \mu_2(t)=0;\\
\text{Case (c): }& \mu_1(t)=\sqrt{2}\sin(2\pi t), \mu_2(t)=\sqrt{2}\cos(2\pi t), \mu_3(t)=0.
\end{align*}
\end{exmp}
The first two cases are simple binary problems. Case (a) corresponds to the situation where $\mS_B \subseteq \mS_W$, and $\beta(t) = \sqrt{2}\sin(2\pi t)$ is the optimal projection for functional LDA.  Case (b) is a typical instance where $\mS_W \perp \mS_B$. The optimal projection is $\beta^*(t) = \sqrt{2}\cos(2\pi t)\in\mS_B$ as perfect classification can be achieved.  Specifically, $\langle \beta^*,X_{1,i}\rangle = 1$ and $\langle \beta^*,X_{2,i}\rangle = 0$ for all $i=1,\ldots, n_k$, where $n_k$ is the number of functions in class $k$. Case (c) combines the situations considered in cases (a) and (b). $\beta\in \mP_W(\mS_B)$ (resp. $\beta^* \in \mP_W^{\perp}(\mS_B)$) can be used to separate the curves in class 2 (resp. 1) from those in the other two classes. Between these two projections, $\beta^*$ is more informative for classification as $\beta^*$ can completely separate the curves in class 2 from those of the other two classes. Specifically, $\langle \beta^*,X_{2,i}\rangle = 1$ and $\langle \beta^*,X_{1,i}\rangle = \langle \beta^*,X_{3,i}\rangle = 0$ for all $i=1,\ldots, n_k$. This example shows that both $\mS_W$ and $\mS_W^\perp$ are helpful for classification and $\beta$ in $\mP_W^{\perp}(\mS_B)$ is more informative.

Since the information in both $\mS_B$ and $\mS_W$ is essential to identify $\beta$'s and generally $\mS_B \nsubseteq \mS_W$, we consider finding $\beta$'s in $\mP_W^{\perp}(\mS_B)$ and in $\mP_W(\mS_B)$ sequentially. We first consider $\beta$'s in $\mP_W^{\perp}(\mS_B)$ because they can lead to asymptotically perfect classification (see Theorem \ref{thm:apc} for details). 
Without loss of generality, we let the global mean $\mu = \sum_{k=1}^c \pi_k \mu_k = 0$, where $\pi_k$ is the probability that a randomly selected function $X$ is from class $k$ and $\sum_{k=1}^c \pi_k=1$. So, the between covariance 
\begin{align*}
\Gamma_B(s,t) & = \sum_{k=1}^c \pi_k \{\mu_k(s)-\mu(s)\}\{\mu_k(t)-\mu(t)\} \\
&= \sum_{k=1}^c \pi_k \mu_k(s)\mu_k(t),
\end{align*}
 for $s,t\in\mathcal{T}$. In practice, $\pi_k$ is unknown and we estimate it with $n_k/\sum_{i=1}^c n_i$. For convenience, we denote $\Gamma \beta = \int_\mathcal{T}\Gamma(s,t)\beta(t)dt$. 

\subsection{sFLDA}
Specifically, sFLDA is defined as finding the optimal projections,  
\begin{align}
& \beta_1 = \argmax_{\beta\in \mP_W^{\perp}(\mS_B), } \langle \beta, \Gamma_B \beta \rangle \text{, and } \nonumber \\
& \beta_j = \argmax_{\beta\in \mP_W^{\perp}(\mS_B), \langle \beta, \beta_i\rangle =0 \text{ for }i<j} \langle \beta, \Gamma_B \beta \rangle \nonumber \\ & \hspace{45mm} \text{for }j=2,\ldots,c'; \label{eq:nuLDAnew} 
\end{align}
\begin{align}
& \beta_{c'+1} = \argmax_{\beta\in \mP_W(\mS_B) } \frac{\langle \beta, \Gamma_B \beta \rangle}{\langle \beta, \Gamma_W \beta \rangle} \text{, and} \nonumber \\
& \beta_{c'+j} = \argmax_{\beta\in \mP_W(\mS_B), \langle \beta, \beta_{c'+i}\rangle =0 \text{ for }i<j} \frac{\langle \beta, \Gamma_B \beta \rangle}{\langle \beta, \Gamma_W \beta \rangle} \nonumber \\ & \hspace{45mm} \text{for } j=2,\ldots, c'', \label{eq:LDAnew}
\end{align}
where $\|\beta\|=1$, $c'$ (resp. $c''$) is the dimension of $\mP_W^{\perp}(\mS_B)$ (resp. $\mP_W(\mS_B)$), and both $c'$ and $c''$ are unknown in practice. These $\beta$'s are optimal in the sense that (i) when $\mS_B\perp\mS_W$, asymptotically perfect classification can be achieved, and (ii) when $\mS_B\subset\mS_W$, they are the optimal projections of functional LDA. To identify $\{\beta_i\}_{i=1}^{c'}$, we introduce a symmetric non-negative definite kernel
\begin{equation}\label{eq:gamma_b}
\Gamma_{B\backslash W}(s,t) = \sum_{k=1}^{c} \pi_k r_k(s) r_k(t),
\end{equation}
where $r_k(t) = \mu_k(t) - \sum_{j=1}^\infty \langle \mu_k, \phi_j \rangle \phi_j(t)$. $r_k$ is the projection of $\mu_k$ on $\mS_W^\perp$ and $r_k\in\mP_W^{\perp}(\mS_B)$. By Mercer's Theorem,  
\begin{equation}\label{eq:gamma_b_eig}
\Gamma_{B\backslash W}(s,t)= \sum_{j=1}^{c'} \eta_j \psi_j(s) \psi_j(t),
\end{equation}
where $\psi_j$ is the $j$th eigenfunction of $\Gamma_{B\backslash W}$ with corresponding eigenvalue $\eta_j>0$ in descending order. Simple calculations (see Appendix \ref{app:detail2.1} for detail) lead to $ \beta_j = \psi_j$ for $j=1,\ldots,c'$, where $c'\leq c-1$. 

Next, we look for $\{\beta_i\}_{i=c'+1}^{c'+c''}$ in  $\mP_W(\mS_B)$. Similar to (\ref{eq:gamma_b})--(\ref{eq:gamma_b_eig}), we define another symmetric non-negative definite kernel
\begin{equation}\label{eq:gamma_b2}
\Gamma_{BW}(s,t) = \sum_{k=1}^{c} \pi_k r^*_k(s) r^*_k(t), 
\end{equation}
where $r_k^*(t) = \mu_k(t) - r_k(t)$. $r^*_k$ is the projection of $\mu_k$ on $\mS_W$, and $r_k^*\in\mP_W(\mS_B)$. Again, by Mercer's Theorem,  
\begin{equation}\label{eq:gamma_b_eig2}
\Gamma_{BW}(s,t)= \sum_{j=1}^{c''} \eta_j^* \psi_j^*(s) \psi_j^*(t),
\end{equation}
where $\eta_j^*>0$ is the $j$th eigenvalue in descending order with corresponding eigenfunction $\psi_j^*(t)$ and  $c'' \leq c-1$.   
Since $\beta(t) = \sum_{i=1}^{c''} a_i\psi_i^* (t)$ for some constant $\bm{a} = (a_1,\ldots,a_{c''})^T$, 
obtaining $\{\beta_j\}_{j=c'+1}^{c'+c''}$ in (\ref{eq:LDAnew}) becomes equivalent to solving the eigenequation 
\begin{equation} \label{eq:flda} 
\Omega_W^{-1}\Omega_B\bm{a}=\zeta\bm{a},
\end{equation}
where $\|\bm{a}\|=1$, $\Omega_B = \text{diag}(\eta_1^*,\ldots,\eta_{c''}^*)$, and the element in $i$th row and $j$th column of $\Omega_W$ is $\langle \psi_j^*, \Gamma_W \psi_j^*\rangle$. The equivalence is detailed in Appendix \ref{app:detail2.1}. Consequently, the noninvertibility issue of $\Gamma_W$ is avoided and finding $\{\beta_i\}_{i=c'+1}^{c'+c''}$ is streamlined to a multivariate problem (\ref{eq:flda}). 

When $\{\beta_i\}_{i=1}^{c'+c''}$ are available, one could apply any classifiers on the projections to perform classification. For illustration purposes and simplicity, we employ the nearest centroid classifier in our analysis.

\subsection{Special Cases}

When $\mS_B \perp \mS_W $, only $\beta_i$'s in (\ref{eq:nuLDAnew}) are considered and asymptotically perfect discrimination can be achieved (shown in Theorem \ref{thm:apc}), 
and case (b) in Example 2.1 is an artificial example with $c=2$.

When $\mS_B \subseteq \mS_W$, only $\beta_{i}$'s in (\ref{eq:LDAnew}) are considered and most existing functional LDA approaches were developed under this specific situation. For example, (2.3) in \cite{DelaigleH:12} implies that $\mS_B \subseteq \mS_W$ is considered, and the authors showed that these $\beta_{i}$'s can lead to asymptotically perfect classification for binary classification problems under some conditions. However, (\ref{eq:flda}) is computationally not only easier but more efficient as the eigenfunctions of $\Gamma_W$ irrelevant to $\mS_B$ are filtered out in (\ref{eq:flda}). A typical example of $\mS_B \subseteq \mS_W$ is the multiplicative random effect model, where the mean function is proportional to one of the eigenfunctions, e.g., \cite{JianAW:09}. Further, Case (a) in Example 2.1 is an artificial example with $c=2$.

Note that when $\mS_B \not\subseteq \mS_W$ and $c=2$, only one $\beta$ in $\mP_W^{\perp}(\mS_B)$ is considered. Specifically, sFLDA identifies $\beta$ via (\ref{eq:nuLDAnew}) with $c'=1$ and asymptotically perfect classification is expected. 


\section{Estimation}
Let $y_{k,ij}$ be the $j$-th observation of subject $i$ in class $k$ made at $t_{k,ij}$, for $j=1,\ldots,m_{k,i}$,  $i=1,\ldots,n_k$, and $k=1,\ldots,c$. Specifically,  
\begin{equation*}
y_{k,ij}  = X_{k,i}(t_{k,ij}) + \epsilon_{k,ij}, 
\end{equation*}
where $X_{k,i}$ is defined as in (\ref{eq:Xkkl}), and $\epsilon_{k,ij}$ is the measurement error with mean zero and variance $\sigma^2$ and is independent from all other random variables. The mean function for class $k$ can be estimated by applying any one dimensional smoother to $\{(y_{k,ij},t_{k,ij})|1\leq j \leq m_{k,i} \text{, } 1\leq i \leq n_k\}$. Take the local linear smoother for example,  
\begin{align}
\label{eq:muk} \hat{\mu}_k(t) = & \hat{b}_0, \text{ where  for } \hat{\bm{b}}=(\hat b_0,\hat b_1),\\
\bm{\hat{b}} = & \arg\min_{\bm{b}}
\sum_{i=1}^{n_k}\frac{1}{m_{k,i}h_k}\sum_{j=1}^{m_{k,i}}
K(\frac{t-t_{k,ij}}{h_k}) \times \{Y_{k,ij}-b_0-b_1(t_{k,ij}-t)\}^2,\nonumber
\end{align} 
$h_k$ is the bandwidth and $K(\cdot)$ is the kernel function. 
The within covariance function $\Gamma_W$  can be estimated by applying a two dimensional smoother to $\{(R_{k,i,j,\ell},t_{k,ij},t_{k,i\ell})| k=1,\ldots,c; i=1,\ldots, n_k; 1\leq j\neq\ell \leq m_{k,i}\},$ where $R_{k,i,j,\ell} = y_{k,ij}^Cy_{k,i\ell}^C$ and $y_{k,ij}^C = y_{k,ij}-\hat\mu_k(t_{k,ij}).$ Take the two dimensional local linear smoother for example,   
\begin{align}
\label{eq:sigma_w} &\hat{\Gamma}_W(s,t)=\hat{b}_0, \text{ where  for } \bm{\hat b}=(\hat b_0,\hat b_1,\hat b_2),\\ \nonumber
&\bm{\hat{b}} = \arg\min_{\bm{b}}
\sum_{k=1}^{c} \sum_{i=1}^{n_k} \frac{1}{m_{k,i}(m_{k,i}-1)h_W^2} \\
& \hspace{5mm} \times \sum_{1\leq j\neq\ell \leq m_{k,i}}
K(\frac{s-t_{k,ij}}{h_W})K(\frac{t-t_{k,i\ell}}{h_W}) \nonumber \\ 
& \hspace{5mm}  \times\{R_{k,i,j,\ell}-b_0-b_1(t_{k,ij}-s)-b_2(t_{k,i\ell}-t)\}^2.\nonumber
\end{align}
The bandwidths of $\hat\mu_k$'s and that of $\hat\Gamma_W$ in our numerical analyses are selected via leave-one-curve-out CV, and the formulae are provided in Appendix \ref{app:bs}. To estimate $\sigma^2$, we employ the approach in \cite{YaoMW:05:1} and denote it as $\hat \sigma^2$. Details are omitted to save space.  We obtain $\hat\lambda_i$'s and $\hat\phi_i$'s by applying an eigendecomposition to $\hat\Gamma_W$. 

\subsection{Estimating $\beta_1,\ldots,\beta_{c'}$}
To estimate $\{\beta_i\}_{i=1}^{c'}$, we suggest performing an eigendecomposition on 
\begin{equation}
\hat{\Gamma}_{B\backslash W}(s,t) = \sum_{k=1}^{c} \frac{n_k}{n} \hat{r}_k(s) \hat{r}_k(t),
\label{eq:gamma_b_est}
\end{equation}
where $\hat{r}_k(t) = \hat{\mu}_k(t) - \sum_{j=1}^L \langle \hat{\mu}_k, \hat{\phi}_j \rangle \hat{\phi}_j(t)$, for $\{\hat{\psi}_i\}_{i=1}^{c'}$ as well as $\{\hat\beta_i\}_{i=1}^{c'}$. 
Both $L$ and $c'$ are selected by fraction of variation explained (FVE). Specifically, 
\begin{equation*}
\begin{split}
&L  = \argmin_{1\leq\ell<\infty } \frac{\sum_{i=1}^\ell \hat\lambda_i}{\sum_{i} \hat\lambda_i} \geq \mathcal{P}_1, \text{ and}\\
&c'  = \argmin_{1\leq\ell\leq c-1} \frac{\sum_{i=1}^\ell \hat\eta_i}{\sum_{i} \hat\eta_i} \geq \mathcal{P}_2,
\end{split}
\end{equation*}
where $\hat\lambda_i$ (resp. $\hat\eta_i$) is the $i$-th eigenvalue of $\hat\Gamma_W$ (resp. $\hat\Gamma_{B\backslash W}$), and $0< \mathcal{P}_1,\mathcal{P}_2 \leq 1$. 

In general, the threshold $\mathcal{P}$ is chosen to be 80\% or 85\% in FPCA; however, we choose 95\% to select both $L$ and $c'$ in our analyses to prevent accidentally excluding the information about $\mS_B$ in $\mP_W^{\perp}(\mS_B)$ or including the information about $\mS_W$ in $\mP_W^{\perp}(\mS_B)$.  Consider a general case where $\mS_B \not\subseteq \mS_W$. Empirically, the estimated $\mS_W$, denoted as $\hat{\mS}_W$, is spanned by $\{\hat\phi_i\}_{i=1}^L$. When the selected $L$ is too large, some unreliable $\hat\phi_i$'s are used to estimate $\mS_W$. Let $\{\hat\phi_i\}_{i=L-d+1}^L$ be these unreliable estimates, where $d$ is some positive integer. If $\{\hat\phi_i\}_{i=L-d+1}^L$ is orthogonal to $\mS_B$, the subsequent classification remains unchanged as $\hat r_k$'s are not affected due to $\langle \hat\mu_k, \hat\phi_j \rangle = O(h_k^2 + \delta_{n_k,1}(h_k))$
for $L-d < j \leq L$ by (\ref{thm:muk}). However, if $\{\hat\phi_i\}_{i=L-d+1}^L$ is not orthogonal to $\mS_B$, which certainly is possible, the subsequent classification tends to be corrupted as some of the information about $\beta$'s in the $\hat r_k$'s might be removed due to $\langle \hat\mu_k, \hat\phi_j \rangle = \langle \mu_k, \hat\phi_j \rangle + O(h_k^2 + \delta_{n_k,1}(h_k))$ and $\langle \mu_k, \hat\phi_j \rangle \not\rightarrow 0$ for $L-d < j \leq L$. When the selected $L$ is too small, $\hat r_k$'s tend not to be orthogonal to $\mS_W$. However, since the number of $\beta$'s can be smaller than the dimension of $\Gamma_{B\backslash W}$, it is possible that $\hat\beta$'s are orthogonal to $\mS_W$. When this happens, asymptotically perfect classification can still be achieved. However, if those $\beta$'s are not orthogonal to $\mS_W$, asymptotically perfect classification might not be achieved as the projections of different classes might not be well separated from each other. So, we need a criterion to select $L$ properly and our empirical experience indicates that FVE is a good choice. Furthermore, FVE is computationally simple and fast and free of model assumptions. 

\subsection{Estimating $\beta_{c'+1},\ldots,\beta_{c'+c''}$}
Similarly to estimating $\{\beta_i\}_{i=1}^{c'}$, we first estimate $\Gamma_{BW}$ by
\begin{equation}
\hat{\Gamma}_{BW}(s,t) = \sum_{k=1}^{c} \frac{n_k}{n} \hat{r}^*_k(s) \hat{r}^*_k(t),
\label{eq:gamma_b_est2}
\end{equation}
where $\hat{r}^*_k(t) = \hat{\mu}_k(t) - \hat r_k(t)$. The estimated eigenfunctions $\hat{\psi}^*_i(t)$'s and estimated eigenvalues $\hat{\eta}^*_i$'s are obtained by applying an eigendecomposition to $\hat{\Gamma}_{BW}$. Again, $c''$ is selected by FVE with threshold $95\%$. \\ $\hat\Omega_B = \text{diag}(\hat{\eta}^*_1,\ldots,\hat{\eta}^*_{c''})$ and the element in $i$th row and $j$th column of $\hat\Omega_W$ is $\langle \hat{\psi}^*_i, \hat\Gamma_W \hat{\psi}^*_j\rangle$. Therefore, $\hat\beta(t) = \sum_{i=1}^{c''} \hat a_i \hat{\psi}^*_i(t)$, where $(\hat a_1,\ldots,\hat a_{c''})^T$ is obtained by solving (\ref{eq:flda}), where $\Omega_W$ and $\Omega_B$ are replaced with $\hat\Omega_W$ and $\hat\Omega_B$, respectively. Given $c'' \leq (c-1)$, letting $c'' = c-1$ may not be particularly detrimental to results; however, it is not necessary to include those $\hat\psi_j^*$'s corresponding to very small $\hat\eta_j^*$'s as they are not reliable when performing classification. That is the main reason that only a truncated number of estimated eigenfunctions are used.

\subsection{Cases where $c'=c-1$}
When $\mS_B \not\perp \mS_W$ and $\mS_B \not\perp \mS_W^\perp$, our procedure estimates the optimal set $\{\beta_j\}_{j=1}^{c'+c''}$ by both (\ref{eq:nuLDAnew}) and (\ref{eq:LDAnew}). However, $\Gamma_{B\backslash W}$ is theoretically zero when $\mS_B \subseteq \mS_W$, but in practice $\hat{\Gamma}_{B\backslash W}$ is a random matrix and has $(c-1)$ non-zero eigenvalues due to random noise. Consequently, when $c'=(c-1)$, the true case ($\mS_B\subseteq \mS_W$ or $\mS_B \perp \mS_W $) needs to be further clarified. Our empirical experience indicates that $q$-fold CV works well. Specifically, we first randomly divide the training sample into $q$ groups. Each time one group is used as the testing sample while the remaining $(q-1)$ groups are applied to perform sFLDA under both cases. The procedure is repeated $q$ times and the decision is made by comparing the overall misclassification rates. The choice of $q$ depends on the sample size, the number of observations per subject and available time for computation. Our experience indicates that $q=5$ is acceptable in our analysis. However, a larger $q$ definitely can help reduce the model misspecification rate and thus the misclassification rate as more samples are used. Please see the supplement for details. 

\subsection{Alogrithm}
The sFLDA procedures can be summarized as Algorithm 1 and the MATLAB code of sFLDA is available at \url{https://github.com/chenlu-hung/SFLDA}.      
\begin{algorithm}\label{alg:1}
\caption{Steps to perform sFLDA.}
\begin{algorithmic}[1]
\REQUIRE
$\{(y_{k,ij},t_{k,ij}): 1\leq i \leq n_k, 1\leq j\leq m_{k,i} \}$ for $k=1,\ldots,c$.
\ENSURE
$\hat\beta_{1},\ldots,\hat\beta_{c'+c''}$.
\STATE Perform an eigendecomposition to $\hat{\Gamma}_{B\backslash W}$ (\ref{eq:gamma_b_est}) to obtain $\hat{\beta}_1,\ldots,\hat{\beta}_{c'}$, where $c'$ is decided by FVE. (\emph{Section 3.1})
\IF {$c'<c-1$ (i.e. $\mS_B \not\perp \mS_W$ and $\mS_B \not\perp \mS_W^\perp$)}
\STATE 
 Perform an eigendecomposition to $\hat{\Gamma}_{BW}$ (\ref{eq:gamma_b_est2}) to obtain $\hat\beta_{c'+1},\ldots,\hat{\beta}_{c'+c''}$, where $c''$ is decided by FVE. (\emph{Section 3.2}) \\
\ELSE 
\STATE Use $q$-fold CV to determine whether $\mS_B \perp \mS_W$ or $\mS_B\subseteq \mS_W$. (\emph{Section 3.3})
	\IF {$\mS_B \perp \mS_W$}
		\STATE Set $c''=0$. 
	\ELSE
		\STATE Set $c'=0$ and 
 		Perform an eigendecomposition to $\hat{\Gamma}_{BW}$ (\ref{eq:gamma_b_est2}) to obtain $\hat\beta_{1},\ldots,\hat{\beta}_{c''}$, where $c''$ is decided by FVE. (\emph{Section 3.2}) \\
	\ENDIF
\ENDIF
\end{algorithmic}
\end{algorithm}

\section{Asymptotics}

Before deriving the theoretical results of $\{\hat\beta_i\}_{i=1}^{c'+c''}$, we list some useful results from \cite{LiH:10:1}. First, we define the $j$th harmonic mean of $m_{k,i}$ as $\gamma_{n_k,j} =  \left( n_k^{-1} \sum_{i=1}^{n_k} 1/m_{k,i}^j \right)^{-1}$,  
\begin{align*}
\delta_{n,1}(h) & = \max_{1\leq k\leq c} \left[\{1+1/(h_k\gamma_{n_k,1})\} \log n_k/n_k\right]^{1/2}\text{, and }\\
\delta_{n,2}(h) & = \max_{1\leq k \leq c} \big[\{1+1/(h_k\gamma_{n_k,1})+ 1/(h_k^2\gamma_{n_k,2}) \} \\
&\hspace{15mm}\log n_k/n_k\big]^{1/2}.
\end{align*}   \cite{LiH:10:1} showed that under Assumptions A.1--A.4, 
\begin{equation}\label{thm:muk}
\sup_{t\in\mathcal{T}} |\hat\mu_k(t)-\mu_k(t)| = O(h_k^2 + \delta_{n_k,1}(h_k)) \text{ a.s.}  
\end{equation}
for $k=1,\ldots,c$ and that under Assumptions A.1--A.6,  
\begin{align}\label{thm:SigWc}
\sup_{s,t\in\mathcal{T}} & |\hat\Gamma_W(s,t)-\Gamma_W(s,t)| \nonumber = O(h^2 + \delta_{n,1}(h) + h_W^2 + \delta_{n2}(h_W)) \text{ a.s.. }
\end{align}  
We assume that $n_k$'s are of the same order, and thus it is reasonable to have $h_k$'s of the same order, $h$. If the order of $h_1$ is smaller or equal to that of $h_2$, we denote it as $h_1 \lessapprox h_2$.

Note that $L$ could be a slowly divergent sequence $L_n$ and similar arguments have been made in \cite{HallH:06} and \cite{HallMW:06}. 
Specifically, if we assume
\begin{equation}\label{A8}
\begin{split}
&\lambda_j > \lambda_{j+1}>0 \text{, } E(A_{k,j}^4)\leq C\lambda_j^2 \text{, and } \\
&\lambda_j-\lambda_{j+1} \geq C^{-1}j^{-(a_1+1)} \text{ for } a_1\geq 1,
\end{split}
\end{equation}
we can provide a sufficient condition of $L$:
\begin{equation}\label{A9}
\text{as }n\rightarrow\infty \text{, } L^{a_1+2}(\delta_{n2}(h_W)+h_W^2) \rightarrow 0,
\end{equation}
where $(\delta_{n2}(h_W)+h_W^2)$ is the $\mL_2$ convergence rate of $\hat\Gamma_W$ (if $h\lessapprox h_W$). A similar argument about $L$ based on similar assumptions can be found in \cite{YaoLW:15}. 


\subsection{Asymptotic Properties of $\hat\beta_1,\ldots,\hat\beta_{c'+c''}$} 
To show the convergence rate of $\hat\beta_1,\ldots,\hat\beta_{c'+c''}$, we need those of $\hat{\Gamma}_{B\backslash W}$, $\hat{\Gamma}_{BW}$, $\hat\Omega_B$ and $\hat\Omega_W$. First, we show the convergence rate of $\hat{\Gamma}_{B\backslash W}$. Due to (\ref{eq:gamma_b_est}), we could instead show the convergence rate of $\hat r_k$, which depends on the decay order of $\langle \mu_k,\phi_i\rangle$ to ensure convergence. So, we assume $\langle \mu_k,\phi_i\rangle \leq D i^{-\alpha}$ for some $D>0$ and $\alpha>1$, and similar arguments can be found in \cite{HallH:07}.  For notation convenience, let $$\Delta_n = L^{2a_1+3}\{h^4 + \delta^2_{n,1}(h) + h_W^4 + \delta^2_{n,2}(h_W) \} +  L^{-(2\alpha-1)}.$$ 

First, we can obtain the following theorem about $\hat r_k$ and $\hat{\Gamma}_{B\backslash W}$.
\begin{thm} \label{thm:rk}
Under Assumptions A.1--A.7,
\begin{equation*}
\begin{split}
&\| \hat r_k - r_k\|^2 = O(\Delta_n)  \text{ a.s. for } k=1,\ldots,c;  \\
&\|\hat{\Gamma}_{B\backslash W} - \Gamma_{B\backslash W}\|^2 = O(\Delta_n)  \text{ a.s.}.
\end{split}
\end{equation*} 
\end{thm}

Similarly, we can have the following theorem about $\hat r^*_k$ and $\hat{\Gamma}_{BW}$.
\begin{thm} \label{thm:rk2}
Under Assumptions A.1--A.7,
\begin{equation*}
\begin{split}
& \| \hat r^*_k - r^*_k\|^2 = O(\Delta_n)  \text{ a.s. for }k=1,\ldots,c;  \\
& \| \hat{\Gamma}_{BW} - \Gamma_{BW}\|^2 = O(\Delta_n)  \text{ a.s..}
\end{split} 
\end{equation*} 
\end{thm}

Simple calculations and Theorem \ref{thm:rk2} lead to  
\begin{thm} \label{thm:Omg}
Under Assumptions A.1--A.7, 
 \begin{align*} 
 &\|\hat\Omega_B-\Omega_B\|^2  = O\left(\Delta_n  \right) \text{ a.s., and} \\ 
& \|\hat\Omega_W-\Omega_W\|^2  = O\left(\Delta_n \right) \text{ a.s..}
 \end{align*}
\end{thm}

The asymptotic properties of $\{\hat\beta_j\}_{j=1}^{c'}$ and $\{\hat\beta_j\}_{j=c'+1}^{c'+c''}$ can be obtained by applying perturbation theory to Theorems \ref{thm:rk} and \ref{thm:Omg}, respectively. Thus, the corollary follows. 
\begin{cor}\label{cor:beta2}
Under Assumptions A.1--A.7 and that nonzero $\eta_j$'s are distinct, for $1\leq j \leq (c'+c'')$, 
\begin{equation}
\sup_{t\in\mathcal{T}}|\hat\beta_j(t) - \beta_j(t)| = O(\sqrt{\Delta_n}) \text{ a.s.. }
\end{equation} 
Under different sampling schemes, for $1\leq j \leq (c'+c'')$,
\begin{itemize}
\item longitudinal data $(i.e., m_{k,i} < \infty),$ \\ 
$\sup_{t\in\mathcal{T}}|\hat\beta_j(t) - \beta_j(t)| = O( L^{a_1+3/2}\{ h^2+ (\frac{\log n}{nh})^{1/2} + h_W^2 + (\frac{\log n}{nh_W})^{1/2}\} + L^{-(\alpha-1/2)}) \text{ a.s. }$
\item functional data $(i.e., m_{k,i} \gtrapprox \frac{1}{h} \rightarrow \infty),$ \\
$\sup_{t\in\mathcal{T}}|\hat\beta_j(t) - \beta_j(t)| = O( L^{a_1+3/2}\{ h^2+ h_W^2 + (\frac{\log n}{n})^{1/2} \} + L^{-(\alpha-1/2)}) \text{ a.s.. }$ 
\end{itemize}
\end{cor}


\subsection{Asymptotically Perfect Discrimination} 

To show the asymptotically perfect classification property, we consider the case where $c''=0$, i.e., all $\beta_i$'s are in $\mP_W^{\perp}(\mS_B)$, for illustration purposes as \cite{DelaigleH:12} has shown that when $\mS_B\subset\mS_W$, asymptotically perfect classification can be achieved for binary problems under certain conditions.  Suppose the function $Y$ to be classified is observed at $(t_1,\ldots,t_m)$ with unknown class label $\kappa$ and $Y$ is further contaminated with measurement error. Specifically, $Y(t_i) = X_\kappa(t_i)+\epsilon$ for $i=1,\ldots,m$, and $\epsilon$ is {\it i.i.d.} measurement error with mean zero and finite variance $\sigma^2$. Denote $\bm\nu_k = (\langle \beta_1, \mu_k \rangle,\ldots, \langle \beta_{c'}, \mu_k \rangle)^T$ for $1\leq k\leq c$ and $\hat{\bm\nu} = (\langle \hat\beta_1, Y \rangle,\ldots, \langle \hat\beta_{c'}, Y \rangle)^T$.

\begin{thm} \label{thm:apc}
 Under conditions listed in Theorem \ref{thm:rk}, we have   
\begin{equation*}
\|\hat{\bm\nu} - {\bm\nu}_\kappa\|^2 =  O\left(\Delta_n+ \frac{\log m}{m}\right) \text{ a.s.}, 
\end{equation*}
and if further $\min_{1\leq i \leq c; i\neq\kappa} \|{\bm\nu}_\kappa- {\bm\nu}_i\|^2 > C \left(\Delta_n +\frac{\log m}{m}\right)$ for some $C>0$,
\begin{equation} \label{eq:kappa}
\kappa = \argmin_{1\leq i \leq c} \|\hat{\bm\nu} - {\bm\nu}_i\| \text{ a.s.}.
\end{equation}
\end{thm}
Theorem \ref{thm:apc} indicates that when all $\beta_i$'s are in $\mP_W^{\perp}(\mS_B)$,  the projection of $Y$, $\hat{\bm\nu}$, will converge to ${\bm\nu}_\kappa$ when $n$ and $m$ are large enough.  Moreover, if ${\bm\nu}_\kappa$ and the other ${\bm\nu}_i$'s are not very close, the class label of $Y$ can be correctly classified by employing any nearest centroid based classifier.

\section{Imputation Approach for Longitudinal Data}\label{reconstruct}

The above mentioned estimators are applicable to both functional and longitudinal data.  With $\{\hat\beta_j\}_{j=1}^{c'+c''}$, having LDA projections for a subject with dense observations for subsequent classification is not difficult. However, the projection is nontrivial for a new subject with sparse observations. One might consider employing the FPCA approach in \cite{YaoMW:05:1} to first reconstruct the curve and perform projections later. However, doing so causes some potential risks. When the magnitude of mean functions is relatively small compared to the first few eigenvalues of $\Gamma_W$ and $\mS_B \perp \mS_W$, the true mean function will never be well preserved through the FPCA reconstruction. Take a binary classification problem for example: for $t\in[0,1]$, $\mu_1(t) = \sin(2\pi t)/10$, $\mu_2(t) = -\sin(2\pi t)/10$, and $\phi_k(t) = \sqrt{2}\cos(2\pi kt)$ and $\lambda_k = 2/k$ for $k=1,\ldots,10$. In the pooled covariance function, $\sin(2\pi t)$ corresponds to the smallest eigenvalue, which is too small to be picked up in practice. Thus, the information about the mean function is lost in the FPCA reconstruction. Therefore, we propose an imputation approach to predict the projections.     

For the projection of a new subject $i$ from {\it unknown} class $k$, we consider 
\begin{equation} \label{eq:cdprj}
E(\langle \beta, X_{k,i} \rangle | \bm{y}^N_{k,i})  =  \langle \beta, E( X_{k,i}  | \bm{y}^N_{k,i}) \rangle, 
\end{equation}
where $\bm{y}_{k,i}^N = (y_{k,i1}^{N},\ldots,y_{k,im_i}^{N})^T$ and \begin{align}\label{eq:reconstr}
E(X_{k,i}|\bm{y}_{k,i}^N) = & \sum_{j=1}^c E(1_{(j=k)}|\bm y_{k,i}^N) \nonumber  \left\{ \mu_{j}(t) + \sum_{\ell=1}^\infty A_{j,i\ell}\phi_\ell(t) \right\}.
\end{align}
The estimators of $\mu_j$ and $\phi_\ell$ have been detailed earlier. Given the class label $j$, the PCS $A_{j,i\ell}$ can be predicted by PACE \citep{YaoMW:05:1} and denoted as $\hat{A}_{j,i\ell}$. We estimate $E(1_{(j=k)}|\bm y_{k,i}^N)$ by a pseudo-likelihood approach, which may seem a little ad hoc; however, it works well in general because it can preserve the mean functions that may not be represented through the FPCA reconstruction, as we mentioned earlier. Specifically, 
\begin{equation}
\hat E(1_{(k=j)}|\bm y_{k,i}^N) =  \frac{(n_j/n)f_j(\bm y_{k,i}^N)}{\sum_{j=1}^c (n_j/n)f_j(\bm y_{j,i}^N)}, 
\end{equation} 
where $f_k(\bm{y}_{j,i}^N) \propto \exp \{-(\bm{y}_{j,i}^N-\bm{\hat\mu}_{k,i})^T\hat\Gamma_{W,k,i}^{-1}(\bm{y}_{j,i}^N-\bm{\hat\mu}_{k,i})\}$, $ \bm{\hat\mu}_{k,i} = \hat\mu_k(\bm T_{k,i})$, $\bm T_{k,i} = (t_{k,i1},\ldots,t_{k,im_i})^T$, and $\hat\Gamma_{W,k,i} = \sum_{\ell=1}^L \hat\lambda_\ell \hat\phi_\ell(\bm T_{k,i})\hat\phi_\ell(\bm T_{k,i})^T + \hat\sigma^2 I_{m_i\times m_i}$. 

To sum up, the projection is predicted by
\begin{align*}
\langle \hat\beta, \hat E(X_{k,i}|\bm{y}_{k,i}^N) \rangle = & \sum_{j=1}^c \hat E(1_{(j=k)}|\bm y_{k,i}^N)   \left\{ \langle \hat\beta, \hat\mu_{j}\rangle + \sum_{\ell=1}^L \hat A_{j,i\ell} \langle\hat\beta,\hat\phi_\ell\rangle \right\}.
\end{align*}



\section{Simulation Studies}\label{sec:sim}
Here we investigate the empirical performance of sFLDA by conducting simulation studies with three different cases on the structure of the mean function. The data is generated from 
\[
y_{k,i}(t)  = \mu_k(t)+\sum_{j=1}^{10}A_{k,i,j} \phi_j(t) + \epsilon, \text{ for } k=1,2,3,
\]
where $\phi_{j}(t)=\sin(2\pi jt)$, $t\in[0,1]$, $A_{k,i,j}\sim N(0,1/j^2)$, and $\epsilon \stackrel{i.i.d.}{\sim} N(0,1/11^2)$. We still consider the same mean structures as follows:
\begin{description}
\item[(a)] $\mu_1(t)=\sin(2\pi t)$, $\mu_2(t)=\sin(4\pi t)$, and $\mu_3(t)=0$;
\item[(b)] $\mu_1(t)=\sin(2\pi t)$, $\mu_2(t)=\sin(2\pi t)+\frac{1}{4}\cos(2\pi t)$, and $\mu_3(t)=0$;
\item[(c)] $\mu_1(t)=\frac{1}{5}\cos(2\pi t)$, $\mu_2(t)=\frac{1}{5}\cos(4\pi t)$, and $\mu_3(t)=0$.
\end{description}  
For each case, we generate 300 random trajectories (100 per $k$) as a training set and an additional 300 random trajectories (100 per $k$) as the testing sample for both functional and longitudinal cases. The functional observations are made on a grid of 200 equispaced points on $[0,1]$ for each subject. For longitudinal data, we randomly select 2 to 10 different observation times from the 200 equispaced points with equal probabilities for each subject. The sFLDA is compared with several widely used methods, including spline-based LDA (FLDA, \cite{JamesH:01}), FPCA+LDA  \citep{Muller:05}, and penalized PLS (PPLS, \cite{KramerBT:08}). Note that the PLS proposed in \cite{DelaigleH:12} is for binary classification and is not directly applicable for a general $c$-category problem. Thus, we compare sFLDA with PPLS instead. The R code for FLDA is adapted from the author's website; the MATLAB package ``PACE'' \citep{YaoMW:05:1} and the R package ``ppls'' \citep{KramerBT:08} are employed to perform FPCA and PPLS, respectively. Each experiment consists of 100 runs. All the tuning parameters of the compared approaches (if any) are selected via leave-one-curve-out CV. 

Next, we elaborate why these three cases are considered. Case (a) considers the situation where $\mS_B \subseteq \mS_W$, in which $\mS_B$ can be well-represented by the first two eigenfunctions, and thus both FPCA+LDA and sFLDA are expected to perform relatively well. In case (b), the mean functions can not be fully represented by the eigenfunctions. This implies that both $\mS_W$ and $\mS_W^\perp$ are informative, but not sufficient, for discrimination. Case (c) is a typical example where $\mS_W \perp \mS_B$. Since the variation between the mean functions is much smaller than the first few eigenvalues, performing FPCA results in the loss of considerable information for discrimination; thus, FPCA+LDA acts similarly to a random guess in this case. 

\begin{table}[htp]
\caption{\label{err_func}Classification error rates (\%) for functional data under three simulation settings.}
\centering
\begin{tabular}{ccccc}
\hline
Case & FLDA & FPCA+LDA   & PPLS  & sFLDA \\
\hline
(a) & $46.1\pm4.0$ & $33.3\pm3.1$   & $53.1\pm3.5$ &  $33.0\pm3.1$ \\
(b) & $42.2\pm4.9$ & $53.5\pm2.6$   & $55.5\pm2.8$ &  $23.3\pm3.0$ \\
(c) & $12.5\pm13.0$ & $66.0\pm3.1$  &  $3.3\pm10.0$ &  $0\pm0.0$ \\
\hline
\end{tabular}
\end{table}

The results of the simulated functional data are summarized in Table \ref{err_func}, indicating sFLDA works very well for all three cases. 
As expected, FPCA+LDA performs similarly to sFLDA and both outperform the other two methods in case (a). In case (b), sFLDA significantly outperforms all the other approaches. In case (c), sFLDA does achieve asymptotically perfect classification as expected. PPLS and FLDA perform much better than FPCA+LDA. As mentioned earlier, FPCA+LDA acts like a random guess as crucial information is lost for discrimination in the FPCA step. 

Although one can always reconstruct longitudinal data, the classification results highly depend on the reconstruction quality and generally are not better than those based on functional data. So, we simply compare sFLDA with the approaches designed for longitudinal data, i.e., FLDA, FPCA+LDA and FPCA+SVM \citep{WuL:13}. The results are summarized in Table \ref{err_long}. FPCA+LDA, FPCA+SVM and sFLDA have similar performance and all outperform FLDA in case (a). sFLDA significantly (resp. slightly) outperforms FPCA+LDA and FPCA+SVM in case (b) (resp. (c)). When the number of observations per subject increases, the performance of sFLDA improves dramatically. However, FPCA+LDA and FPCA+SVM do not perform significantly better with the increase in $m_{k,i}$. Please refer to Table 3 in the supplement and Table \ref{err_func} for more details. Generally, FLDA does not perform well in all three cases. Comparing Table \ref{err_long} with Table \ref{err_func}, FLDA, FPCA+LDA, and sFLDA perform similarly or worse due to fewer observations. 
Table \ref{err_long} also provides the error rates under correctly specified scenario (Oracle) and the number of incorrect decisions made by $q$-fold CV. As expected, $q$-fold CV does not perform well in case (b) due to the mean functions only being partially represented by the eigenfunctions of $\Gamma_W$ while additional useful information is contained in $\mS_W^\perp$. This complex model structure makes the model selection quite challenging, especially for sparsely and irregularly observed longitudinal data. However, $q=5$ appears to work nicely as the sFLDA misclassification rates are very close to those under Oracle.     

\begin{table*}[htp]
\caption{\label{err_long}Classification error rates (\%) for longitudinal data under three simulation settings, where M/M stands for model misspecification rate out of 100 runs due to performing $q$-fold CV.}
\centering
\begin{tabular}{ccccccc}
\hline
Case & FLDA & FPCA+LDA & FPCA+SVM & sFLDA & M/M & Oracle\\
\hline
(a) & $44.9\pm2.9$ & $38.4\pm2.7$ & $38.6\pm2.7$ & $37.5\pm2.8$ & 25 & $37.0\pm2.8$ \\
(b) & $55.6\pm2.6$ & $57.3\pm3.7$ & $57.1\pm3.6$ & $46.1\pm3.3$ & 83 & $46.2\pm3.9$ \\
(c) & $59.3\pm3.9$ & $60.7\pm4.7$ & $60.1\pm4.8$ & $54.0\pm5.0$ & 5 & $54.4\pm4.5$ \\
\hline
\end{tabular}
\end{table*}

\section{Data Analysis} Two real data examples under different configurations are considered. For the functional dataset, we compare sFLDA with FLDA, FPCA+LDA and PPLS. For the longitudinal dataset, sFLDA is compared with FLDA, FPCA+LDA, FPCA+SVM and PPLS. As PPLS is not designed for longitudinal data, we reconstruct the latent trajectories by the imputation approach in Section \ref{reconstruct} and perform PPLS to the reconstructed curves. All the tuning parameters for the existing approaches (if any) are selected by leave-one-curve-out CV.  

\subsection{Functional Data} The phoneme dataset (available at \url{http://statweb.stanford.edu/~tibs/ElemStatLearn/}) is used here. The dataset consists of 4509 speech frames (transformed into log-periodogram of length 256) of five phonemes (872 frames for ``she'', 757 frames for ``dark'', 1163 frames for the vowel in ``she'', 695 frames for the vowel in ``dark'', and 1022 frames for the first vowel in ``water''). 
To evaluate the performance of different approaches, we split the dataset into training and testing sets 100 times. In each split, we randomly select $n$ log-periodogram samples per phoneme for training, and the remaining ones are for testing. The misclassification rates of different approaches with different training sample size $n$ are summarized in Table \ref{Phoneme_results}, indicating sFLDA outperforms all the other approaches, and PPLS works better than FLDA and FPCA+LDA. Our algorithm selects three LDA directions from (\ref{eq:nuLDAnew}) and one from (\ref{eq:LDAnew}). This suggests that $\mS_B \not\subseteq \mS_W$ may be more suitable for this data. This dataset further demonstrates the advantage of our approach for multi-category classification, where the LDA directions may be in a combination of (\ref{eq:nuLDAnew}) and (\ref{eq:LDAnew}).

\begin{table}
\caption{\label{Phoneme_results}Misclassification rates (mean$\pm$std\%) of Phoneme dataset.}
\centering
\begin{tabular}{rcccc}
\hline
$n$ & FLDA & FPCA+LDA & PPLS & sFLDA \\
\hline
50 &  $13.8\pm0.6$ & $16.3\pm0.7$ &  $10.0\pm0.5$ & $9.0\pm0.5$ \\
100 &  $11.8\pm0.4$  & $16.3\pm0.6$ & $9.7\pm0.5$ & $7.8\pm0.5$ \\
\hline
\end{tabular}
\end{table}

\subsection{Longitudinal Data}
\begin{figure}
\centering
\includegraphics[width=0.7\linewidth]{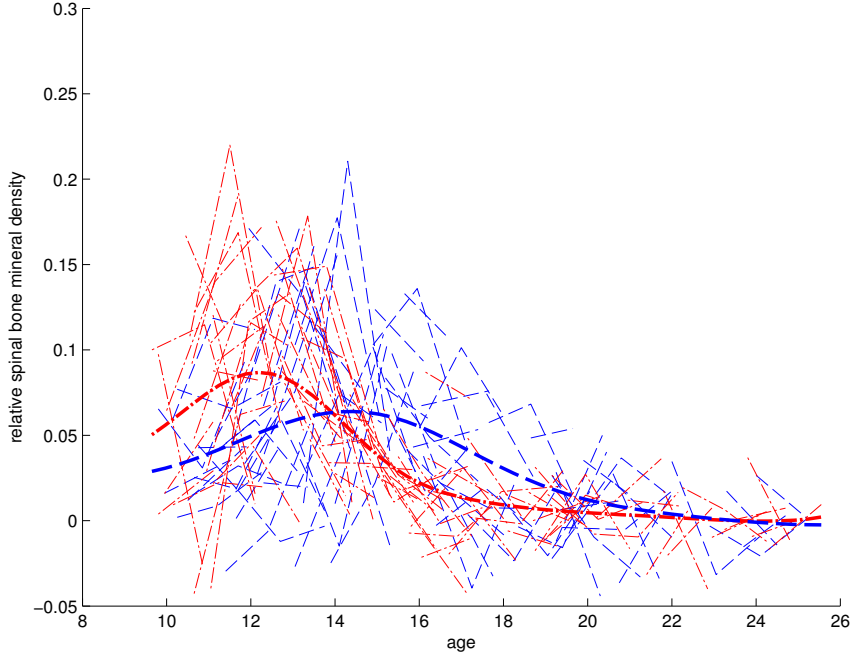}
\caption{Bone densities measured on 154 North American adolescents (blue dot-dashed: male; red dashed: female).}
\label{bone_raw}
\end{figure}

The relative spinal bone mineral density dataset (\cite{BachrachHWNM:99}, available at \url{http://statweb.stanford.edu/~tibs/ElemStatLearn}) is considered. The measurements were made on $154$ North American adolescents with $70$ male and $84$ female children. The observation $y_{k,i,j}$ represents the relative spinal bone mineral density for child $i$ measured at age $t_{k,i,j}$. The measured densities are shown in Figure \ref{bone_raw}, with females in red (dot-dashed) and males in blue (dashed). Even though females and males have different development patterns (e.g. females develop earlier than males), the development also varies from subject to subject, which makes the classification difficult. 

The leave-one-out misclassification rates of sFLDA, FPCA+LDA, FPCA+SVM, PPLS and FLDA are $29.2\%$, $30.1\%$, $30.1\%$, $33.1\%$ and $35.7\%$, respectively. The first three approaches perform similarly and all slightly outperform FLDA and PPLS. FPCA+LDA works better than PPLS, which suggests that the scenario $\mS_B \subseteq \mS_W$ is more appropriate for this case. Our algorithm selects the LDA direction through (\ref{eq:LDAnew}) and has the lowest misclassification rate.



\section{Conclusions}

We have proposed sFLDA for both functional data and longitudinal data to find the optimal LDA projections $\beta$'s in $\mP_W^{\perp}(\mS_B)$ and $\mP_W(\mS_B)$ sequentially. Theoretically, one could follow the technique in \cite{HeMW:03} to tackle the noninvertibility issue of $\Gamma_W$ while extending LDA to functional data directly. However, our strategy through (\ref{eq:flda}) is more appealing in that not only is the noninvertibility issue avoided but it is computationally easier. We have also investigated the asymptotic properties of the proposed estimators. When all the $\beta$'s are in $\mP_W^{\perp}(\mS_B)$, we have shown that sFLDA can achieve asymptotically perfect discrimination when a nearest centroid classifier is applied to the projected data. 

The framework of sFLDA was developed under the LDA settings, where the covariance functions among classes are identical. When the covariance structures among groups are different, a few functional approaches based on the idea of quadratic discriminant analysis have been proposed, such as \cite{JamesH:01}, and \cite{DelaigleH:12}. Extending sFLDA for such general cases requires a more sophisticated procedure as the space spanned by the eigenfunctions becomes much more complicated. We have been working on this general problem with a completely different strategy and this remains an interesting direction for future work.

Although sFLDA originates from extending Fisher's LDA to its functional version, it also works well empirically on high dimensional (HD) multivariate data classification (please see the supplement for details). Note that (\ref{eq:muk}) and (\ref{eq:sigma_w}) can simply be replaced with other empirical estimates as no smoothing is needed in HD data. SVM is one of the best classification approaches and is also the most widely used for HD data classification; however, it requires lots of computational effort due to its complex quadratic computational algorithm and the need to select tuning parameters. The computational burden becomes serious as the data dimension and the sample size increases, which has particular relevance in the big data era. Our numerical investigations have shown that sFLDA with less computational cost still yields comparable performance with SVM, especially when the sample size is moderately large. Therefore, the proposed approach seems quite competitive and promising in this era of big data. 

\section*{Acknowledgements}
The authors wish to thank Professor John A.D. Aston at Cambridge University for his insightful comments and careful proofreading which improved the presentation of the paper substantially.

\appendix
\section{Assumptions}\label{app:assumption}

Since the estimators $\hat\mu_k(t)$ and $\hat\Gamma_W(s,t)$ are estimated by local linear smoothers, it is natural to make the standard smoothness assumptions on the second derivatives of $\mu_k$ and $\Gamma_W$. It is assumed that in class $k$, the data $(\mathbf{T}_i,\mathbf{Y}_{k,i}), i=1,\cdots, n_k,$ have the same distribution, where $\mathbf{T}_{k,i}=(T_{k,i1},\cdots,T_{k,im_{k,i}})$ and $\mathbf{Y}_{k,i}=(Y_{k,i1},\cdots,Y_{k,im_{k,i}})$.  Notice that $(T_{k,ij},Y_{k,ij})$ and $(T_{k,i\ell},Y_{k,i\ell})$ are dependent but identically distributed. Assume the density of time at observation to be  $g(t)$.  Suppose $Y_{k,ij} = \mu_k(T_{k,ij})+U_{k,ij}$, where cov$(U_i(s),U_i(t))=\Gamma_W(s,t)+\sigma^2I(s=t)$ and $\Gamma_W(s,t)=\sum_{\ell=1}^\infty \lambda_\ell \phi_\ell(s)\phi_\ell(t)$.  Additional assumptions and conditions are listed below and similar ones can be found in \cite{LiH:10:1}. 
\begin{itemize}
\item[A.1] For some constant $\Delta_t>0$ and $\Delta_T<\infty$, $\Delta_t\leq g(t) \leq \Delta_T$ for all $t\in\mathcal{T}$. Further, $g(\cdot)$ is differentiable with a bounded derivative.
\item[A.2] The kernel function $K(\cdot)$ is a symmetric probability density function on $[-1,1]$ and is of bounded variation on $[-1,1]$. Further, we denote $\nu_2 = \int_{-1}^1 u^2K(u)du$.
\item[A.3] The mean function $\mu_k$'s are twice differentiable and their second derivatives are bounded.
\item[A.4] $E(|U_{k,ij}|^{\lambda}) < \infty$ and $E(\sup_{t\in\mathcal{T}} |X_k(t)|^{\delta}) < \infty$ for some $\delta \in (2,\infty)$; $h_k \rightarrow 0$ and $(h^2_{k}/\gamma_{n1})^{-1} (\log n_k/n_k)^{1-2/\delta} \rightarrow 0$ as $n_k \rightarrow \infty$.
\item[A.5] All second-order partial derivatives of $\Gamma_W$ exist and are bounded on $\mathcal{T}\times\mathcal{T}$.
\item[A.6] $E(|U_{k,ij}|^{2\delta_\phi}) < \infty$ and $E(\sup_{t\in\mathcal{T}} |X_k(t)|^{2\delta_\phi}) < \infty$ for some $\delta_\phi \in (2,\infty)$; $h_W \rightarrow 0$ and \\$(h^4_W+h^3_W/\gamma_{n1} + h^2_W/\gamma_{n2})^{-1}(\log n/n)^{1-2/\delta_\phi} \rightarrow 0$ as $n \rightarrow \infty$.
\item[A.7] $\langle \mu_k,\phi_i\rangle \leq D i^{-\alpha}$ for some positive constant $D$, where $\alpha>1$.
\end{itemize}

\section{Bandwidth Selection} \label{app:bs}
The bandwidths of $\hat\mu_k$'s and that of $\hat\Gamma_W$ are chosen via
leave-one-curve-out CV as suggested by \cite{RiceS:91}. Specifically, 
\begin{equation*} \label{eq:mukcv} h_k=\arg\min_{h\in \mathbb{R}^+} \sum_{j=1}^{n_k}
 \frac{1}{m_{k,j}}\sum_{\ell =1}^{m_{k,j}}\left\{Y_{k,j\ell}-\hat{\mu}_k^{(-j)}(t_{k,j\ell})\right\}^2,
\end{equation*}
where $\hat{\mu}_k^{(-j)}(t_{k,j\ell})$ is the estimated $\mu_k(t_{k,j\ell})$ when $h$ is the bandwidth and the observations of the $j$-th curve are not used to estimate $\mu_k$. Similarly, the bandwidth for $\Gamma_W$ is defined as
\begin{equation*} \label{eq:GWcv} 
\begin{split}
h_W  = & \arg\min_{h\in \mathbb{R}^+}  \sum_{k=1}^{c} \sum_{j=1}^{n_k}  \frac{1}{m_{k,j}(m_{k,j}-1)} \sum_{1\leq \ell_1 \neq\ell_2 \leq m_{k,j}} \left\{R_{k,j,\ell_1,\ell_2}-\hat{\Gamma}_W^{(-j)}(t_{k,j\ell_1}, t_{k,j\ell_2})\right\}^2,
\end{split}
\end{equation*}
where $\hat{\Gamma}_W^{(-j)}(t_{k,j\ell_1}, t_{k,j\ell_2})$ is the estimated $\Gamma_W(t_{k,j\ell_1}, t_{k,j\ell_2})$ when $h$ is the bandwidth and $R_{k,j,\ell_1,\ell_2}$'s of the $j$-th curve are not used to estimate $\Gamma_W$.

\section{Some Details for Section 2.1}\label{app:detail2.1}
\subsection{$\{\beta_i\}_{i=1}^{c'} \in \mP_W^{\perp}(\mS_B)$}
To show: $\beta_j = \psi_j$ for $j=1,\ldots,c'$ \\
Note that $\mu_k(t)= r_k(t)+ r^*_k(t)$ for $t\in\mathcal{T}$, where $r_k\in\mP_W^{\perp}(\mS_B)$ and $r^*_k(t) = \sum_{j=1}^\infty \langle \mu_k, \phi_j \rangle \phi_j(t)$ is in $\mP_W(\mS_B)$. Simple calculations lead to $\Gamma_B = \Gamma_{B \backslash W} + \mathcal R_e$,
where
\begin{align*}
&\Gamma_B(s,t)  = \sum_{k=1}^c \pi_k \mu_k(s)\mu_k(t)\text{, and } \\
&\Gamma_{B\backslash W}(s,t)  = \sum_{k=1}^c \pi_k r_k(s)r_k(t).
\end{align*}
Clearly,  $\langle \beta, \Gamma_B \beta \rangle = \langle \beta, \Gamma_{B \backslash W} \beta \rangle$ if $\beta \in \mP_W^{\perp}(\mS_B)$. Therefore, $\beta_j = \psi_j$. 

\subsection{$\{\beta_i\}_{i=c'+1}^{c'+c''}\in\mP_W(\mS_B)$}
To show: $\beta(t) = \sum_{i=1}^{c''} a_i\psi_i^* (t)$ \\
Since $\mP_W(\mS_B)$ is the space spanned by $\{r^*_k\}_{k=1}^{c}$ and  $\psi_i^*$'s are the eigenfunctions of $\Gamma_{BW}$, a given $\beta\in \mP_W(\mS_B)$ can be represented as $\beta(t) = \sum_{i=1}^{c''} a_i\psi_i^* (t)$, where $a_i$'s are basis coefficients. Direct calculations lead to 
\begin{equation}\label{eq:flda_numerator} 
\langle \beta, \Gamma_{BW} \beta \rangle= 
\bm{a}^T\Omega_B\bm{a},
\end{equation}
and
\begin{equation}\label{eq:flda_denominator} 
\langle \beta, \Gamma_W \beta \rangle= 
\bm{a}^T\Omega_W\bm{a}.
\end{equation} 
Combining (\ref{eq:flda_numerator}) and (\ref{eq:flda_denominator}), (\ref{eq:LDAnew}) becomes
\begin{equation}\label{eq:flda_gev} 
\begin{split}
\max_{\beta\in \mP_W(\mS_B), \|\beta\|=1 } \frac{\langle \beta, \Gamma_B \beta \rangle}{\langle \beta, \Gamma_W \beta \rangle}&=\max_{\beta=\sum_{i=1}^{c''} a_i\psi_i^*, \|\beta\|=1} \frac{\langle \beta, \Gamma_B \beta \rangle}{\langle \beta, \Gamma_W \beta \rangle} \\
&=\max_{\|\bm{a}\|=1}\frac{\bm{a}^T\Omega_B\bm{a}}{\bm{a}^T\Omega_W\bm{a}},
\end{split}
\end{equation}
because $\langle \beta, \Gamma_B \beta \rangle = \langle \beta, \Gamma_{BW} \beta \rangle$ when $\beta \in \mP_W(\mS_B)$, and $\|\beta\|=1$ implies $\|\bm a\|=1$. Therefore, (\ref{eq:LDAnew}) is equivalent to (\ref{eq:flda_gev}). Since $\mP_W(\mS_B) \subseteq \mS_W$, $\Omega_W$ is nonsingular and thus invertible. As a result, (\ref{eq:flda_gev}) is equivalent to (\ref{eq:flda}). Once $\bm{a}$ is solved by (\ref{eq:flda}), we have $\beta(t) = \sum_{i=1}^{c''} a_i\psi_i^* (t)$.

\bibliography{fda}
\bibliographystyle{chicago}

\end{document}